\documentclass{article}
\usepackage{graphicx} 
\usepackage[round]{natbib}
\usepackage{amsmath, amssymb}
\usepackage{booktabs}
\usepackage{multirow}
\usepackage{booktabs}
\usepackage{xcolor}
\usepackage{enumitem}
\usepackage{rotating}
\usepackage{ragged2e} 

\usepackage{geometry}
\providecommand{\keywords}[1]{\textbf{{Keywords---}} #1}

\title{Bridging the Gap: Introducing Joint Models for Longitudinal and Time-to-event Data in the Social Sciences}
\author{Sophie Potts, Anja Rappl, Karin Kurz, Elisabeth Bergherr}
\date{\today}

\begin{document}

\maketitle

\begin{abstract}
    \noindent 
    In time-to-event analyses in social sciences, there often exist endogenous time-varying variables, where the event status is correlated with the trajectory of the covariate itself. Ignoring this endogeneity will result in biased estimates. In the field of biostatistics this issue is tackled by estimating a joint model for longitudinal and time-to-event data as it handles endogenous covariates properly. This method is underused in the social sciences even though it is very useful to model longitudinal and time-to-event processes appropriately. Therefore, this paper provides a gentle introduction to the method of joint models and highlights its advantages for social science research questions. We demonstrate its usage on an example on marital satisfaction and marriage dissolution and compare the results with classical approaches such as a time-to-event model with a time-varying covariate. In addition to demonstrating the method, our results contribute to the understanding of the relationship between marriage satisfaction, marriage dissolution and other covariates.
\end{abstract}

\noindent \keywords{joint models, longitudinal data, time-to-event data, marriage dissolution, relationship satisfaction}

\section{Introduction}

Research questions pointing to the risk of experiencing an event as well as respective data sets are frequently found in social science research, e.g.\ in family formation \citep{kurz2006case, Kingsley2018}, educational attainment \citep{Ameri2016}, recidivism \citep{Skardhamar2012} or reemployment \citep{Hgglund2017}.
They are commonly estimated with hazard models from the field of time-to-event analysis. These research questions, including the examples from above, often involve time-varying covariates (TVC) which allow to model the impact of covariates that change over time.
The classical approach to include the TVCs in time-to-event models is based on the last value carried forward (LVCF) strategy (from now on referred to as \emph{TVC approach}). This allows to make individual-specific predictions for each time point during their individual observation period but makes the assumption, that the value does not change between two observation times.
Furthermore, this strategy only holds appropriate estimation results, when the TVC is exogenous, i.e.\ is independent of the time-to-event outcome (event happened vs.\ censored). 
The exogeneity assumption does not hold in cases of anticipatory effects of the event or when the trajectory of the TVC is highly correlated with the outcome (more on the definition of exogeneity and endogeneity in Section \ref{sec: whynot}).
Both assumptions, LVCF and exogeneity, are particularly questionable for many frequently changing and self-reported covariates as they are common in social sciences when individual trajectories of TVC and person-related events are of interest.

In these cases a joint models for longitudinal and time-to-event data \citep{Wulfsohn1997} is an appropriate estimation routine. It combines a longitudinal estimation procedure for the TVC and a classical time-to-event model. Including the estimates of the longitudinal model in the time-to-event model links them to a joint model. The estimation of the regressions coefficients of the two linked submodels is carried out simultaneously. 
It therefore allows to model the relationship between an endogenous covariate and the risk of an event appropriately and represents a useful tool to investigate complex social research questions.

Joint models are standard tools in biostatistics, e.g.\ to investigate the relationship between the trajectory of a biomarker in blood cells on all-cause mortality \citep{Nez2014} or on recurrence of cancer \citep{Ferrer2016} but do not yet belong to the standard toolkit of social science researchers \citep{Cremers.2021}. In order to increase the usage in social science applications a low threshold introduction to the method is needed. Therefore, this tutorial paper aims to guide the reader through a social science application of a joint model.
As an illustrative application, we will model the marital satisfaction and the risk of marriage dissolution using a German panel data base \citep{huinink2011pairfam}. 


The following section highlights the definition of endogeneity and exogeneity in order to lay the foundation to identify endogenous covariates properly. Section \ref{sec: JM} introduces the method of joint models for longitudinal and time-to-event data and its properties using the example of marital satisfaction and the risk of marriage dissolution. 
After a short review of the literature of marital satisfaction and dissolution, Section \ref{sec: marsat} describes the data base, the model specification for the application and discusses the estimation results. In Section \ref{sec: comparison} the results are compared to other modelling strategies (classical TVC approach, two-stage model). The tutorial concludes with a summary and a discussion of possible extensions as well as problems of the application. The example is executed using the Software \texttt{R} and can be reproduced using the web supplementary material of this tutorial.

\section{Types of time-varying covariates}
\label{sec: whynot}

Regarding the question whether a classical TVC approach models the data appropriately, the type of the time-varying covariate is crucial. The usage of the TVC approach in time-to-event models does not pose problems, when the modelled covariate is exogenous. In contrast, endogenous variables are problematic in the case of classical hazard models as they do result in biased estimates and therefore do not allow for causal statements strictly speaking \citep{Box-Steffensmeier2004}. Thus, researchers should consider using a joint model when exogeneity is questionable.
Therefore, we shortly review the definition of exogeneity and endogeneity in time-to-event models before describing the method. 

\cite{Kalbfleisch2002} divide exogenous time-varying variables into two sub-categories: defined and ancillary TVCs. \emph{Defined} ones have a predetermined path in advance for all subjects of the study, e.g.\ historical period, cohort or age of the individual. 
In contrast, an \emph{ancillary} TVC is the result of a stochastic process, which is \emph{external to the observation unit} such as population level characteristics, e.g.\ unemployment rate in an economy \citep{Yamaguchi1991}.
However, such exogenous covariates may be rather rare in micro-level analyses in social science research, since many time-varying covariates describe individual-specific changes.

Endogeneous time-varying variables are also categorized into two different sub-types \citep{Kalbfleisch2002}: \emph{state dependent} and \emph{rate dependent} TVCs. The former comprises variables whose path is not independent of the state of the outcome variable. Consequently, they result in different paths of the TVC depending on the respective time-to-event outcome (e.g.\ marital satisfaction trajectories of still married vs.\ separated). "In other words, the value of the time-dependent covariate carries information about the state of the dependent process." \citep[p.~132]{Blossfeld2001} 

Rate dependent covariates are directly correlated with the hazard rate of an event such that not only the trajectory of the TVC correlates with the outcome but the estimated risk of having an event influences the trajectory as well. This can be for example due to anticipation of the event, e.g.\ the effect of the anticipation of divorce on the working behaviour of women \citep{poortman2005}.

\section{Joint models for longitudinal and time-to-event data}
\label{sec: JM}

In this section, we start with an illustrative example and point out the shortcomings of classical time-to-event models when dealing with endogenous time-varying covariates. This is followed by an explanation of the method of joint models for longitudinal and time-to-event data and their advantages.

\subsection{An illustrative example: Time-to-event model and joint model in comparison}

\begin{figure}[ht!]
    \centering
    \includegraphics[width=\textwidth]{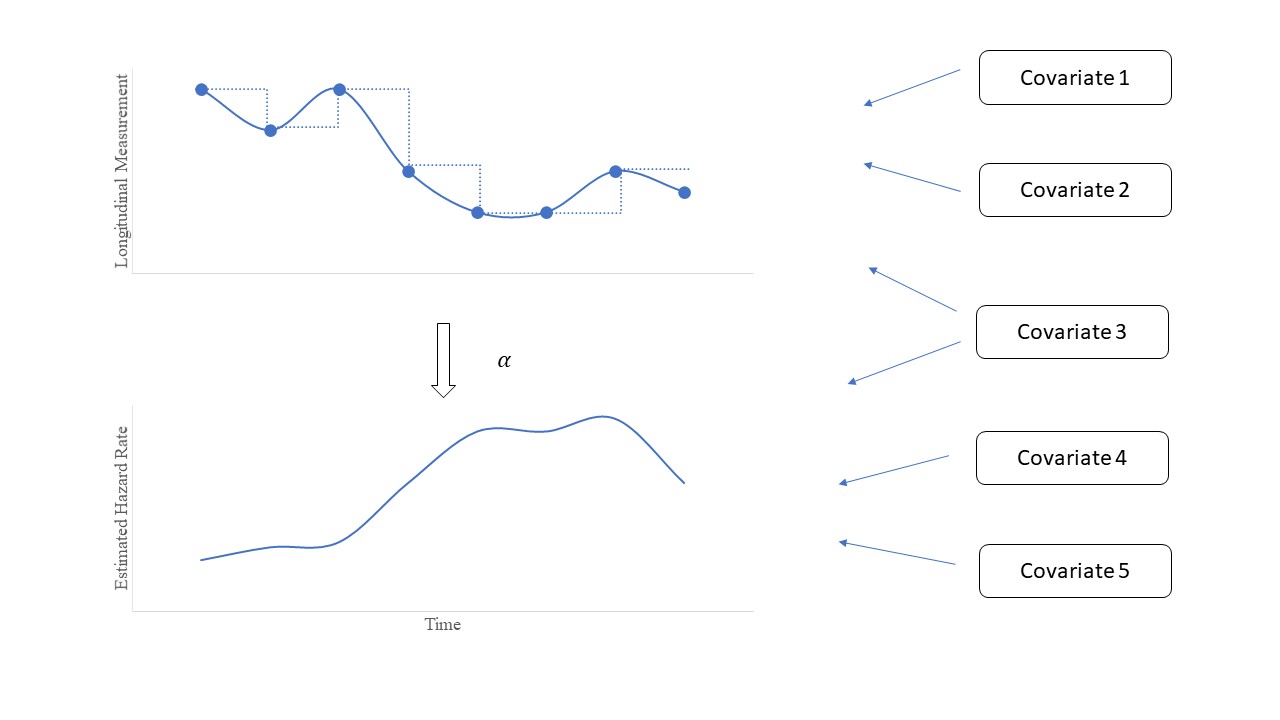}
    \caption{Scheme of two related processes and possible modelling strategies. Upper panel: Measurements of the time-varying covariate (longitudinal process). Lower panel: Risk of having an event (time-to-event process). In a joint model the influence of the trajectory in the upper panel on the risk is estimated by an association parameter $\alpha$ and both processes can be modelled as functions of (shared) covariates.}
    \label{fig:scheme}
\end{figure}

In the example we are interested in - the risk of marriage dissolution (event) and how covariates influence this risk - we may include covariates that are changing over time such as subjective marital satisfaction (TVC). Research interest lies in the influence of the trajectory of marital satisfaction and the risk of marriage dissolution. Assume marital satisfaction has been captured multiple times over the years and information on the start and possible end of the marriage are available. Figure \ref{fig:scheme} serves as an illustration for the fictional example using one single individual: the upper trajectory depicts his/her marital satisfaction measure, where the points correspond to the measurements in time. The measurement points are connected via the smooth trajectory function (solid line). The lower part represents the estimated hazard rate for marriage dissolution over time. In this example higher values of satisfaction (upper panel) go hand in hand with smaller estimated hazard rates, i.e.\ lower risk of ending the relationship (lower panel).

%


In order to include the time-varying covariate marital satisfaction into a time-to-event model, one could use the classical TVC approach. 
This approach yields several problems since it assumes that the respective variable (1) does not change between the observation times and (2) is exogenous. These two assumptions are particularly questionable for volatile and self-reported variables. The first assumption would result in a step-function (dashed line in upper panel of Figure \ref{fig:scheme}) between the measurements in the upper panel instead of the smooth path. Especially for infrequently measured variables with long periods between two measures this approach may model the trajectory inappropriately.

\begin{figure}[h!]
    \centering
    \includegraphics[width=0.9\textwidth]{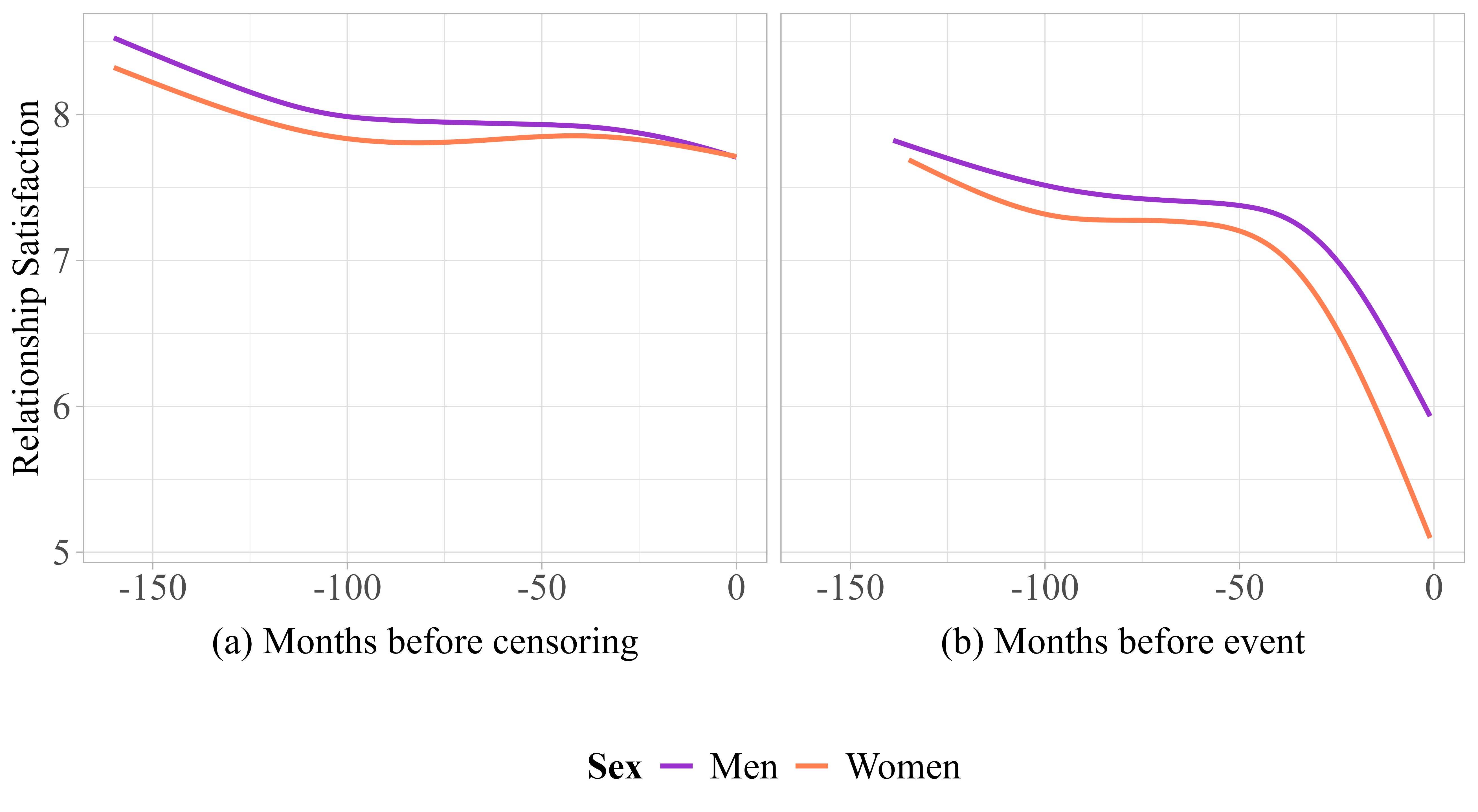}
    \caption{Estimated average trajectory of relationship satisfaction of persons by event status and sex. Non-linear smoother by sex (dark: men, light: women). Based on the illustration by \cite{crowther2013_stjm}.}
    \label{fig:summary_trajectories}
\end{figure}

Additionally, as marital satisfaction does neither evolve from a stochastic process, which is external to the individual under study (ancillary TVC), nor can be calculated as a defined covariate, it can be called an endogenous TVC and therefore violates the second assumption. Figure \ref{fig:summary_trajectories} may be an indicator for state dependence of marital satisfaction showing the smoothed average trajectory of marital satisfaction clustered by marital status: (a) persons still in the relationship and (b) persons that ended their relationship to their married partner. The fact that the trajectories (both, for men and women) differ significantly between the marital status groups, is a strong indicator for a state dependent TVC and thus a modelling technique able to appropriately adress this endogeneity has to be applied.
Arguably rate dependence may also apply in this example, as \cite{clark2008lags} found a strong anticipatory effect of the life event \emph{divorce} regarding life satisfaction for men and women. It seems plausible that a similar effect exists for marital satisfaction.

Besides the proper estimation in the presence of endogeneous time-varying covariates, an additional difference between the joint model and the classical time-to-event model is the inclusion of predictors for modelling the trajectory of the TVC. The basic approach of a TVC does not allow to investigate its predictors (Figure \ref{fig:scheme}: Covariate 1, 2, 3 would not be taken into account for the upper panel).

Even though a two-stage approach can be applied to first model the TVC as a function of covariates in order to include their effects and further overcome the problem of LVCF, it has some unfavourable statistical properties mainly arising from the independent estimation of the two submodels. 
The subsequent time-to-event model treats the prediction of the firstly fitted longitudinal outcome, as if it was estimated without any uncertainty. This results in an underestimation of standard errors for the TVC and thus cannot be interpreted appropriately in terms of statistical inference.

\noindent Compared to these two approaches the joint model for longitudinal and time-to-event data is advantageous as it

\begin{itemize}
    \item can handle endogenous TVC in time-to-event models
    \item allows for proper inference in the presence of endogenous covariates due to simultaneous estimation routine
    \item allows for changes in the TVC between observation points
    \item deals with informative drop-out in longitudinal studies (see Section \ref{sec: perspect}).
\end{itemize}
A main drawback of joint models is the computational effort during the fitting procedure. In contrast, two-stage models are less computationally demanding. Therefore, some research focuses on combining the advantages of the two methods -- the unbiased estimates from the joint model and the fast estimation of the two-stage approach \citep{LeivaYamaguchi2020}.

\subsection{Method}

Joint models \citep{FAUCETT1996, Wulfsohn1997, rizopoulos.2012} overcome the above mentioned shortcomings as they allow for joint modelling of a repeatedly measured outcome alongside the risk of having an event of interest. Rather than using the observations of the TVC, the joint model considers the repeatedly measured TVC as the result of a longitudinal process subject to its own model (marital satisfaction, Figure \ref{fig:scheme}: upper part). This longitudinal process is combined with the related time-to-event process (risk of marriage dissolution, Figure \ref{fig:scheme}: lower part). 
The two submodels are described for the example in Figure \ref{fig:scheme} before merging them to the joint model. For a more general notation we refer the reader to \cite{Hickey2016}. 

First, the time-varying covariate is modeled via an appropriate model, in general a linear mixed model (LMM)\footnote{An introduction to linear mixed models can be found in \cite{Gaecki2012}.}, allowing for intra-individual variance along the time axis captured by random intercepts ($b_{0i}$) and possibly random slopes $b_{1i}$ (commonly used on time as a covariate).
The model corresponding to Figure \ref{fig:scheme} can be written as

\begin{equation}
\label{eq: longi_general}
\begin{aligned}
    y_{i}(t) &= \underbrace{\left (\beta_0 + b_{0i}\right) + \left ( \beta_1 + b_{1i} \right)t + \beta_2 x_{i1} +\beta_3 x_{i2} + \beta_4 x_{i3} (t)}_{m_i(t)}+ \varepsilon_{i}(t)\\
\end{aligned}
\end{equation}
with $\varepsilon_{i}(t) \sim \mathcal{N}(0, \sigma^2)$ where $i$ indicates the individual and $t$ is the time point of the measurement.
In vectorized form the model can be rewritten as
\begin{equation}
\label{eq: longi_general2}
\begin{aligned}
    {y}_i(t)&= \boldsymbol{x}_{i_{\text{long}} }(t)'\boldsymbol{\beta}+ \boldsymbol{z}_i(t)'\boldsymbol{b}_i+ {\varepsilon}_i(t)
\end{aligned}
\end{equation}
where $\boldsymbol{x}_{i_{\text{long}} }(t)'$ is a row vector with all covariate values and a leading 1 for the intercept for person $i$ at time $t$ and $\boldsymbol{z}_i(t)' = (1 ~~t)'$ holds the covariate values of the random effects, in this case a random intercept and a random slope on time $t$. Using this model allows to incorporate covariates as predictors of the estimated values of the TVC $y_i(t)$. The covariates may be time-constant or time-varying  with $\boldsymbol \beta$ being the regression coefficient vector. As in a classical LMM, the random intercepts and random slopes are assumed to stem from a multivariate normal distribution $\boldsymbol{b}_i \sim \mathcal{N}(\boldsymbol{0}, \boldsymbol{Q})$. 

The second related process is a time-to-event model\footnote{Several classes of time-to-event models are explained by \cite{Blossfeld2001}.}, which is used to model the risk of having an event over time. The general form is a proportional hazards model, which consists of a baseline hazard $h_0(t)$ scaled by a covariate part. The corresponding equation to Figure \ref{fig:scheme} is given by
\begin{equation}
\label{eq: surv general}
\begin{aligned}
    h_i(t) &= h_0(t) \exp[\gamma_0 + \gamma_1 x_{i3}(t) + \gamma_2 x_{i4}(t) + \gamma_3 x_{i5}].
\end{aligned}
\end{equation}
It can be rewritten in vectorized form as 
\begin{equation}
\label{eq: surv general}
\begin{aligned}
 h_i(t) &= h_0(t) \exp [ \boldsymbol{x_{i_{\text{surv}}}}(t)' \boldsymbol \gamma].
\end{aligned}
\end{equation}
The modeled hazard function $h_i(t)$ states the instantaneous risk of person $i$ of having an event at time $t$ (i.e.\ have a marriage dissolution). This model also contains covariates (Figure \ref{fig:scheme}: Covariates 3--5), which may be exogenous time-varying or time-constant and a vector of coefficients $\boldsymbol \gamma$.

In order to take marital satisfaction as an \emph{endogeneous} covariate into the model, a joint model links the estimated value of the TVC process ${m}_{i}(t)$ to the time-to-event model by incorporating it as a predictor and thus estimates their coefficients jointly: 
\begin{equation}
    \begin{aligned}
    \label{eq: jm general}
        h(t|M_i(t), \boldsymbol{x_i}) = h_0(t) \exp [\boldsymbol{{x}_i}_{\text{surv}}(t)' \boldsymbol \gamma + \alpha m_i(t)].
    \end{aligned}
\end{equation}
The coefficient $\alpha$ is called the \emph{association parameter}.
In contrast to the two-stage approach all coefficients ($\beta$, $b$, $\gamma$, $\alpha$) are estimated simultaneously such that all uncertainty is included in the estimation procedure. Estimating the value of marital satisfaction $\hat{m}_{i}(t)$ involves all available observations of the person, such that the hazard in a joint model at this time does implicitly also depend on the covariate history $M_i(t)$.



By including a covariate in both submodels (e.g. $x_{i3}(t)$ in Equation \eqref{eq: jm general}) its direct and indirect effect on the risk of having an event can be separated. Thinking of a variable which has a strong influence on the TVC and further a smaller but significant impact on the survival: By estimating a unique $\hat \beta$ coefficient as well as a $ \hat \gamma$ coefficient and the association parameter $ \hat \alpha$ we decompose the total effect via: $\hat \alpha \hat \beta + \hat \gamma$. Hereby $\hat \alpha \hat \beta$ indicates the mediated effect of the covariate via the trajectory and $\hat \gamma$ represents the direct effect on the risk of having an event. This decomposition is especially helpful to understand the effect pathway of the respective covariate.

Estimation of the coefficients can be done using a Maximum Likelihood approach (Ex\-pec\-ta\-tion-Maximization Algorithm) or Bayesian Methods (Markov-Chain Monte-Carlo sampling). The different estimation strategies for joint models are presented and compared by \cite{Rappl2021}.
Joint models for longitudinal and time-to-event data are implemented in several R packages \citep{jmpackage, joineRpackage} and are also available in STATA \citep{crowther2013_stjm, Crowther2020_merlin}. 
Furthermore, joint models allow for individual-specific predictions as they control for individual characteristics via the random effects in the longitudinal model.

\subsection{Perspectives and extensions of joint models}
\label{sec: perspect}

In the section before, we presented the joint model with a focus on the time-to-event submodel. However, there is also literature focusing on the longitudinal submodel where the time-to-event submodel is used to model informative dropout from the longitudinal study \citep{Asar2015, HOGAN1997, Vonesh2005, wu1988estimation}. This strand of literature refers to joint models as \emph{shared parameter models}. 

In terms of extensions of the presented joint model, a first option is the linkage between the longitudinal and the time-to-event model, as there exist several options of \textit{association structures}. Using the current value of the longitudinal model $m_i(t)$ (as presented in our main joint model in Equation \eqref{eq: jm general}) associates the predicted value at each time-point with the hazard function at the same time-point. One can also think of the slope of the estimated trajectory of the TVC (marital satisfaction) to be important for the hazard function. An increase or decrease in marital satisfaction independent of the actual level may influence the risk of marriage dissolution.
The options can be combined.
Note that the effect size of the slope association structure depends on the units of time, whereas the current value does not. Thus, the estimates of the association structures cannot be compared in their effect size.
Both options can be used in lagged versions as well. Another common approach for the association structure is the usage of estimated random effects of the longitudinal submodel of each person and link them to their survival. The interpretation of the respective association parameter does not depend on time, since random intercept and random slope do not depend on time by default. For an overview of association structures see \cite{Cremers.2021}.

The basic concept of joint modelling can easily be extended to more complex model structures, such as different types of time-varying covariates (e.g.\ count, categorical), competing risks time-to-event models, multiple longitudinal models with correlation, semiparametric modelling of effects using splines and delayed entries. 
For an extensive overview on recent developments in the joint modelling literature see \cite{Papageorgiou2019}.

\section{Application: Marriage satisfaction and time to marriage dissolution}
\label{sec: marsat}

In order to demonstrate the use of joint models in sociology, the relationship between satisfaction with the marriage and the time to marriage dissolution is investigated.
There is a huge body of literature in the field of marital satisfaction, predictors of marriage dissolution/divorce and interrelations of the two. Some studies focus on questions of general development of marital satisfaction throughout the marriage \citep[e.g.][]{Lorber2014, Williamson2019}, others investigate predictors for marital satisfaction \citep[e.g.][]{Elmslie2014, Huss2019}. Some include marital satisfaction as a mediator between the risk of marriage dissolution and other effects in regression models, e.g.\ personality traits \citep{Solomon2014} or household work \citep{Frisco2003}. There are cross-sectional and longitudinal studies, with different degrees of exploitation of the longitudinal structure (two time-points vs.\ whole trajectory) of the data. Different methods were applied to investigate the effect of marital satisfaction on marriage dissolution. The latent growth curve approach of \cite{Lorber2014} for example indicates that the trajectory of marital satisfaction throughout the period of marriage should be modelled individual-specific. Most empirical studies fit separate models to male and female respondents, since the determinants of marriage dissolution and marital satisfaction differ between sexes. 
For a review of theoretical models regarding marital satisfaction evolution see \cite{caughlin2006}. Following the large body of literature, we chose the most common covariates for marital satisfaction; our selection largely matches the findings of the meta analysis of \cite{Karney1995}.

We would like to highlight the paper of \cite{Frisco2003} as it analyses the relationship between the two outcomes of interest in a regression. Their focus is to determine the influence on household work (in)equity on the odds of divorce and the possible mediating effect of marital satisfaction. Without considering individual trajectories, they find a small mediating effect of marital satisfaction but still state a significant direct positive effect of unfair high workload of household work on the odds of divorce eight years later for women. The study is based on measurements at two points in time.

To the best of our knowledge, so far no one has used a joint model for longitudinal and time-to-event analysis to investigate the relationship between marital satisfaction and time to marriage dissolution yet. As this model type allows to exploit the whole richness of data, i.e. the longitudinal character of the data as well as the information of timing of an event, we believe that it is highly suitable to generate more well-founded answers to the questions What determines marital satisfaction?, What determines marriage dissolution? as well as How does marital satisfaction mediate influences on the hazard of marriage dissolution? with respect to the joint evolution of both processes over time.

\subsection{Data set}
The analysis is based on the German pairfam data \citep{pairfamdata}. Pairfam ("Panel Analysis of Intimate Relationships and Family Dynamics") is a longitudinal study with 14 annual waves contributing to shed light to changes in family and relationship structures. It started in 2008 with over 12,000 respondents. Another sample of 1,489 East-German anchor persons ("DemoDiff") is merged as a supplementary to the data base. A detailed description of the pairfam study can be found in \cite{huinink2011pairfam}. The relationship biographies of the respondents as well as the annual questionnaire about the satisfaction with the relationship can be used to build a joint model. 

The final sample consists of all persons in the pairfam data, who were married in their first marriage over the course of at least three interviews. 
This restriction has been made due to two reasons: First, some measurements of the longitudinal variable on marriage satisfaction are needed for proper analysis. Second, using only the first marriage of a person is based on previous empirical findings that relationship stability differs between the first and following marriages and that a selection bias may be present \citep{Jensen2016}, which might also skew the results of the model.
This leaves us with a final sample size of $N=3616$ persons/marriages of which $247$ ($\approx 7\%$) stated an end of this relationship during the observation period (number of events). We did not take the actual month of divorce as event time but the stated end of relationship (marriage dissolution). 

\begin{figure}[h!]
    \centering
    \includegraphics[width=0.75\textwidth]{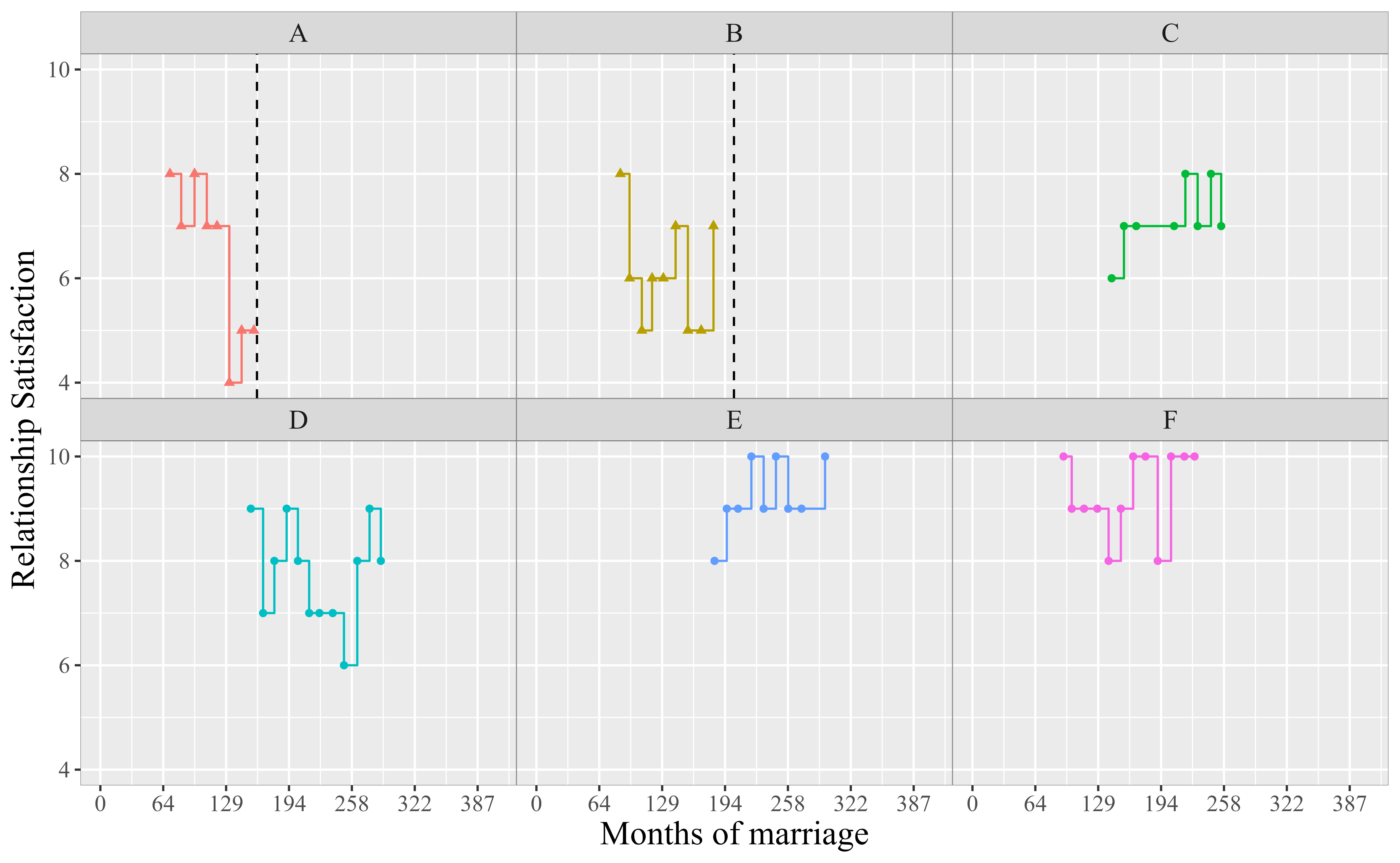}
    \caption{Example trajectories of relationship satisfaction from pairfam participants. Vertical dashed lines indicate the time of marriage dissolution.}
    \label{fig:trajectories}
\end{figure}

Marital satisfaction is measured as the answer on an 11-point scale to the question "All in all, how satisfied are you with your relationship?". Some exemplary trajectories of marital satisfaction are depicted in Figure \ref{fig:trajectories}. Note that marriages which started before the first interview such as the person in panel C in Figure \ref{fig:trajectories}, are not left censored for the time-to-event model since we know the start of their marriage. They just start at a different point in time with time-dependent information. In order to allow users to reproduce the analysis, a synthesized data set can be found in the web supplementary material\footnote{The data set has been generated using the \texttt{simPop} package \citep{simpop}.}.

\begin{table}[h!]
\caption{Description of the final sample. For the covariates means (medians) of non-standardized time-constant variables are presented.}
\vspace{0.1cm}
\label{tab: datatable}
\centering
\begin{tabular*}{\linewidth}{@{\extracolsep{\fill}} l|c|c }
\toprule
& \multicolumn{2}{c}{Descriptive statistics}\\
 & Women & Men \\
\midrule
Number of persons &2,141&1,475\\
Number of events & 161 (7.5\%) & 86 (5.8\%)\\
Number of observations per person & 7.6 (6) & 7.5 (7)  \\
Age at marriage & 27.0 (26) & 29.2 (29)  \\
Relationship duration at marriage in months & 64.2 (54) & 64.9 (54) \\
\multirow{2}{*}{Premarital cohabitation} & Yes: 1800 (84\%)& Yes: 1235 (84\%)\\
 & No:~ 341 (16\%) & No:~ 240 (16\%) \\
\bottomrule
\end{tabular*}
\end{table}

Table \ref{tab: datatable} summarises the time-constant variables used from the pairfam data that are included in the models. 
Relationship duration at marriage, age at marriage and premarital cohabitation were included as time-constant covariates and years of education, personal net income, amount of household work, labor force status, children (presence of preschool child and number of children under 18 in the respondents household) and gender role attitudes are included as time-dependent exogenous variables. The variable on the division of household work is a weighted sum index resulting from five survey items. It is measured on a 5-point scale with endpoints indicating that the respondent does all the work (high values) or the partner does all household work (low values) and thus is a relative measure between the spouses. Further, gender role attitudes is also a sum index over three items (see Appendix \ref{append: index_building household} and \ref{append: index_building gender}). 
Note that all metric variables were z-score standardized for model estimation purposes and the scale of the time is changed from months to vary in the interval $[0,1]$ with 1 being the overall latest time-point measured in a marriage in the sample.

\subsection{Model specification}

Based on the reviewed literature on marital satisfaction and marital dissolution, two joint models will be fitted separately for men and women. They are identical in terms of included covariates, method and association structure.

The \emph{longitudinal model} on satisfaction with the marriage will be modelled by an LMM including a random intercept and a random slope term for the duration of marriage ($t$). 
We included a random slope since intercepts and slopes differ between individuals (see Figure \ref{fig:trajectories} for some example trajectories).
The above mentioned variables are taken into account as covariates to model the longitudinal variable properly. Hereby, $i$ is an indicator for the person and $j$ denotes the measurement at time-point $j$:

\begin{equation}
\label{eq:longi_divorce}
\begin{aligned}
    \hat m_{ij} &= \left (\hat \beta_0 + \hat b_{0i}\right) + \left ( \hat \beta_1 + \hat b_{1i} \right)t_{ij} + \hat \beta_2~ t_{ij}^2 +  \\ &~ \hat \beta_3~ \texttt{age at marriage}_{i} +  \hat \beta_4~ \texttt{relationship duration at marriage}_{i} +\\ &~ \hat \beta_{5}~ \texttt{premarital cohabitation}_{i} + \hat \beta_6~ \texttt{years of education}_{i} +\\ &~ \hat \beta_7~ \texttt{net income}_{ij} +  \hat \beta_8~ \texttt{shared work}_{ij} + \\&~ \hat \beta_{9}~ \texttt{gender attitudes}_{ij}+ \hat \beta_{10}~ \texttt{preschoolchild}_{ij}+ \\ &~   \hat \beta_{11}~ \texttt{nchild=1}_{ij}+ \hat \beta_{12}~ \texttt{nchild=2}_{ij}+  \hat \beta_{13}~ \texttt{nchild=more}_{ij}+   \\ &~ \hat \beta_{14}~ \texttt{lfs=not working}_{ij}+ \hat \beta_{15}~ \texttt{lfs=part-time}_{ij}+ \hat \beta_{16}~ \texttt{lfs=other}_{ij}.
\end{aligned}
\end{equation}
The \emph{time-to-event model} on time to marriage dissolution will be modelled jointly with the longitudinal model from Equation \eqref{eq:longi_divorce}, where the estimated values of satisfaction explain the risk of marriage dissolution. Specifically, the time-to-event model includes a B-Spline approximation of the baseline hazard $h_0(t)$.

Equation \eqref{eq: jm pairfam} shows the final representation of our main joint model on marital satisfaction and the risk of marriage dissolution.
\begin{equation}
\label{eq: jm pairfam}
    \begin{aligned}
        \hat h(t|M_i(t), {\boldsymbol{x}_i}_{\text{surv}}) &= h_0(t) \exp [\hat \gamma_1 \texttt{premarital cohabitation}_{i} +  \\ & \hat \gamma_2~ \texttt{age at marriage}_{i} +  \hat \gamma_3~ \texttt{relationship duration at marriage}_{i} + \\ & \hat \gamma_4~ \texttt{net income}_{i}(t) + \hat \gamma_5~ \texttt{years of education}_{i} + \hat \gamma_6~ \texttt{shared work}_{i}(t) + \\& \hat \gamma_{7}~ \texttt{gender attitudes}_{i}(t) + \hat \gamma_8~ \texttt{preschoolchild}_{i}(t)+ \\ &  \hat \gamma_{9}~ \texttt{nchild=1}_{i}(t)+ \hat \gamma_{10}~ \texttt{nchild=2}_{ij}+  \hat \gamma_{11}~ \texttt{nchild=more}_{i}(t)+   \\ & \hat \gamma_{12}~ \texttt{lfs=not working}_{i}(t)+ \hat \gamma_{13}~ \texttt{lfs=part-time}_{i}(t)+ \hat \gamma_{14}~ \texttt{lfs=other}_{i}(t) + \\ 
        &~\hat \alpha (\hat m_{i}(t))]
    \end{aligned}
\end{equation}

\subsection{Implementation}

For implementation of the above model we use the package \texttt{JM} \citep{jmpackage} for \texttt{R} (version 4.4.0) \citep{Rversion}. \texttt{JM} combines two models that are built with their specific \texttt{R} packages.
The longitudinal model is constructed using \texttt{lme()} from the \texttt{nlme} package \citep{nlme} and therefore requires the usual data structure in long format, where each individual spans several rows corresponding to the observation time points, each holding the covariate value of the time point, respectively.

The time-to-event model is fitted using \texttt{coxph()} from \texttt{survival} \citep{survival-package}. The structure of the underlying data set is equivalent to the long format start-stop-event logic when further exogenous TVCs are used, i.e.\ several rows per individual, indicating the current measured values. In case of lack of other exogenous TVCs in the model, the data set for the time-to-event model reduces to one row per individual and supplements the long format data set for the longitudinal submodel (see \cite{rizopoulos.2012}). Since our model includes other exogenous TVCs (e.g.\ labor force status) one single data set in the classical start-stop-event logic is used.

Both models serve as inputs for the final \texttt{jointModel()} command. For further information on the (optional) arguments in the \texttt{jointModel()} function, we refer to \cite{jmpackage}. \texttt{R} code and a synthesized data set to replicate the example can be found in the web appendix of this paper.

\begin{verbatim}
# time-to-event model
modsurv_female <- coxph(Surv(time = t1, time2 = t2, event = status)~ 
    yeduc + ageatm + preschoolchild + nchild + premarcohab +
    sw_weight + incnet + relduratmar + lfs_rec + genderatt_s, 
    data = df_female, x = TRUE, model=T, cluster = id)

# longitudinal model
modlong_female <- lme(sat31 ~ t + I(t^2) + sw_weight + ageatm + 
    preschoolchild + nchild + premarcohab + yeduc + incnet +
    relduratmar + lfs_rec + genderatt_s, 
    data = df_female, random = ~ t | id)

# joint model for longitudinal and time-to-event data
modjoint_female <- jointModel(modlong_female, modsurv_female, 
    timeVar = "t", method = "spline-PH-GH", 
    control = list(verbose=T, iter.EM=100))
\end{verbatim}

\subsection{Estimation Results}

Separate models were fitted for men and women. This section starts with the estimation results of the joint model for \emph{women}.

The \emph{longitudinal submodel} estimates the relationship between the covariates and marital satisfaction (outcome) (Table \ref{tab: res women} left side). The model results in a U-shaped effect of time. 
The number of years of education shows a positive, significant effect on marital satisfaction. 
Some other covariates show negative, linear effects on satisfaction with the relationship: higher age at marriage and higher personal net income are associated with a significantly lower marital satisfaction. Women with children (compared to childless women) also reveal lower values of satisfaction. There is even an additional negative effect if there are preschool children present in the household. The index for the division of household work reveals a negative and statistically significant effect, i.e.\ women who stated to do more household work are less satisfied with their relationship during marriage.
The relationship duration at time of marriage and the gender role attitudes show insignificant coefficients in this model.

\begin{table}[ht]
\caption{Model for \emph{women}: Regression coefficients of the joint model for longitudinal and time-to-event data.}
\label{tab: res women}
\resizebox{\textwidth}{!}{
\small
\centering
\begin{tabular}{lrrr|rrr}
  \hline
       & \multicolumn{3}{c}{Longitudinal submodel} &  \multicolumn{3}{c}{Time-to-event submodel}\\
  \vspace{0.1cm}
     & \multicolumn{3}{c}{(Marital satisfaction)} &  \multicolumn{3}{c}{(Risk of marriage dissolution)}\\
     \vspace{0.1cm}
  Variable & Estimate & Std. err.  & p-value & Estimate & Std. err. & p-value \\ 
  \hline
(Intercept) & 8.7403 & 0.1054& 0.0000& & &  \\ 
Time& -3.0387&  0.4023 & 0.0000& & &  \\ 
Time$^2$& 2.0987 & 0.5022 & 0.0000& & &  \\ 
Relative load of household work&  -0.1369 &  0.0191 &  0.0000  &  0.1089 &  0.0793 &  0.1694   \\ 
Premarital cohabitation$^a$: yes&  -0.1036 &  0.0788 &  0.1885  &  0.1297 &  0.2225 &  0.5599   \\ 
Age at marriage&  -0.1820 &  0.0361 &  0.0000  &  -0.1546 &  0.1070 &  0.1483   \\ 
Preschool child(ren) in hh$^a$: yes&  -0.0693 &  0.0412 &  0.0924  &  -0.3286 &  0.2121 &  0.1213   \\ 
Number of children in hh$^b$: 1&  -0.2520 &  0.0648 &  0.0001  &  0.0893 &  0.2969 &  0.7637   \\ 
Number of children in hh$^b$: 2&  -0.2594 &  0.0729 &  0.0004  &  0.0508 &  0.3082 &  0.8691   \\ 
Number of children in hh$^b$: more&  -0.2182 &  0.0921 &  0.0178  &  0.0799 &  0.3649 &  0.8266   \\ 
Years of education&  0.0672 &  0.0310 &  0.0299  &  -0.2228 &  0.0969 &  0.0214   \\ 
Personal net income&  -0.0457 &  0.0248 &  0.0651  &  -0.0230 &  0.1462 &  0.8750   \\ 
Relationship duration at marriage&  0.0183 &  0.0340 &  0.5910  &  -0.2570 &  0.1047 &  0.0140   \\ 
Labor force status$^c$: not working&  0.0474 &  0.0625 &  0.4479  &  -0.3514 &  0.3051 &  0.2494   \\ 
Labor force status$^c$: other&  -0.0577 &  0.0677 &  0.3941  &  -0.0238 &  0.2830 &  0.9329   \\ 
Labor force status$^c$: part-time employed&  -0.0170 &  0.0555 &  0.7594  &  -0.1381 &  0.2422 &  0.5685   \\ 
Gender role attitudes&  -0.0082 &  0.0219 &  0.7069  &  0.0819 &  0.0896 &  0.3606   \\ 
Satisfaction ($\hat \alpha$)& & & &  -0.5552 &  0.0551 &  0.0000   \\ 

   \hline
\end{tabular}
}
\small{Reference categories:
$^a$ no, 
$^b$ zero,
$^c$ full-time employed
}
\end{table}
\noindent Next, we examine the \emph{time-to-event submodel} for women (Table \ref{tab: res women} right side) including the properly modelled endogenous variable marital satisfaction. This model indicates which variables still have a direct effect on the risk of marital dissolution when controlling for marital satisfaction.
Looking at the association parameter (last row in Table \ref{tab: res women}), we observe the expected strong negative effect of marital satisfaction on the risk of marriage dissolution: the higher the current value of satisfaction with the relationship, the lower the risk of marriage dissolution. 
Besides this relationship, there are only few significant effects in the time-to-event submodel. Higher educated women and women that were in a long relationship with their married partner before marriage have a lower risk of marriage dissolution. As an example for the decomposition of effects, we focus on the variable of relative household work in the following. In contrast to other research results \citep[e.g.][]{Frisco2003}\footnote{Note, that the measures of household tasks differ, as \cite{Frisco2003} use a measure of feeling of fairness, whereas our variable measures to which extend a person does more or less of the household work. Furthermore, they concentrate on dual-earner households with data from the United States whereas our data base contains first marriages in Germany without any restrictions on the labor force status.}, the relative load of household work done by a person has no direct effect for women on their risk of marriage dissolution. There still remains the indirect effect via the mediator of marital satisfaction: a higher share of household work done by the female respondent results in a significantly lower marital satisfaction which results in significantly higher risk of marriage dissolution. The estimated total effect of the index on the divison of household work on the risk of marriage dissolution adds up to $-0.5552 \times -0.1369 + 0.1089 \approx 0.1849$.

For \emph{men}, the results reveal interesting differences in both submodels (Table \ref{tab: res men}):
Regarding the marital satisfaction (\emph{longitudinal model}), socio-economic variables such as education as well as personal net income do not show significant effects. Similarly to the model for women, children decrease the relationship satisfaction in the marriage compared to childless men and the effect size is larger than for women. In contrast to the model for women, there is no additional significant, negative effect of preschool children. 
Even though the labor force status shows an influence on the marital satisfaction for men, these results have to be interpreted with caution, as over 80\% of the person periods for men indicate a full-time employment.
The relative load of household work has a smaller effect on marital satisfaction and is also negative and statistically significant. 

\begin{table}[ht]
\caption{Model for \emph{men}: Regression coefficients of the joint model for longitudinal and time-to-event data.}
\label{tab: res men}
\resizebox{\textwidth}{!}{
\small
\centering
\begin{tabular}{lrrr|rrr}
  \hline
       & \multicolumn{3}{c}{Longitudinal submodel} &  \multicolumn{3}{c}{Time-to-event submodel}\\
  \vspace{0.1cm}
     & \multicolumn{3}{c}{(Marital satisfaction)} &  \multicolumn{3}{c}{(Risk of marriage dissolution)}\\
     \vspace{0.1cm}
  Variable & Estimate & Std. err.  & p-value & Estimate & Std. err. & p-value \\ 
  \hline
(Intercept) &  9.0393 &  0.1233 &  0.0000  & & &  \\ 
Time&  -2.8529 &  0.4976 &  0.0000  & & &  \\ 
Time$^2$&  1.5018 &  0.7095 &  0.0343  & & &  \\ 
Relative load of household work&  -0.0603 &  0.0267 &  0.0237  &  0.2413 &  0.1312 &  0.0660   \\ 
Premarital cohabitation$^a$: yes&  -0.2385 &  0.1052 &  0.0234  &  -0.3357 &  0.3020 &  0.2663   \\ 
Age at marriage&  -0.1686 &  0.0417 &  0.0001  &  0.1291 &  0.1383 &  0.3508   \\ 
Preschool child(ren) in hh$^a$: yes&  0.0249 &  0.0507 &  0.6238  &  0.0196 &  0.2806 &  0.9442   \\ 
Number of children in hh$^b$: 1&  -0.3302 &  0.0790 &  0.0000  &  -0.1158 &  0.4041 &  0.7744   \\ 
Number of children in hh$^b$: 2&  -0.3910 &  0.0901 &  0.0000  &  0.3430 &  0.3825 &  0.3699   \\ 
Number of children in hh$^b$: more&  -0.3562 &  0.1152 &  0.0020  &  0.3927 &  0.4633 &  0.3967   \\ 
Years of education&  0.0341 &  0.0393 &  0.3864  &  -0.1266 &  0.1232 &  0.3043   \\ 
Personal net income&  0.0017 &  0.0192 &  0.9307  &  -0.0481 &  0.1647 &  0.7703   \\ 
Relationship duration at marriage&  -0.0015 &  0.0420 &  0.9708  &  -0.0709 &  0.1208 &  0.5573   \\ 
Labor force status$^c$: not working&  -0.2574 &  0.0977 &  0.0085  &  0.0613 &  0.4931 &  0.9012   \\ 
Labor force status$^c$: other&  -0.3296 &  0.0927 &  0.0004  &  -0.0242 &  0.3599 &  0.9465   \\ 
Labor force status$^c$: part-time employed&  -0.2288 &  0.1191 &  0.0548  &  -1.1292 &  1.0137 &  0.2653   \\ 
Gender role attitudes&  0.0281 &  0.0263 &  0.2846  &  0.1294 &  0.1162 &  0.2654  \\ 
Satisfaction ($\hat \alpha$)& & & & -0.4534& 0.0702& 0.0000 \\

   \hline
\end{tabular}
}
\small{Reference categories:
$^a$ no, 
$^b$ zero,
$^c$ full-time employed
}
\end{table}

\newpage
\noindent The estimated association parameter $\hat \alpha$ in the \emph{time-to-event submodel} is also significant but the effect size is smaller compared to the model for women. In other words, the estimated marital satisfaction is not as predictive for the risk of marriage dissolution for men as it is for women.
Another difference between the sexes in the submodel regarding the risk of marriage dissolution is the effect of the division of household work: This covariate has a positive and statistically significant effect on the risk of marriage dissolution in the model on men ($p = 0.0660$). In contrast to the model for women, the indirect mediated effect is supplemented by a direct effect of household work on the risk of marriage dissolution. The estimated total effect of this predictor variable on the risk of marriage dissolution can be derived as $-0.4534 \times -0.0603 + 0.2413 \approx 0.2686$.
\begin{figure}[h!]
    \centering
    \includegraphics[width=\textwidth]{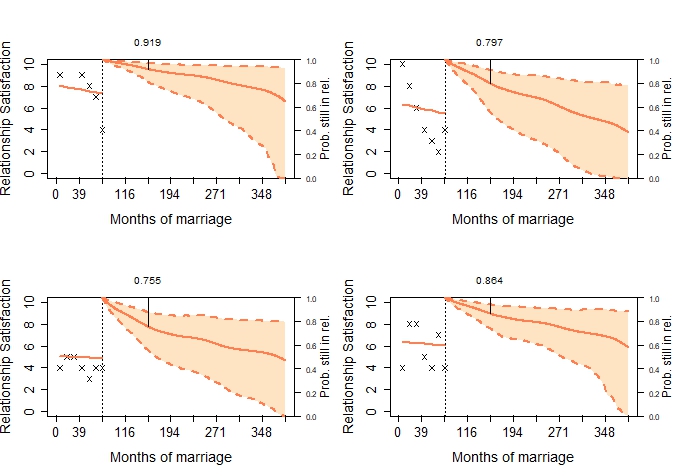}
    \caption{Predicted probability of still being in the marriage for a person varying only the marital satisfaction trajectory. Covariate values: woman, part-time working, 2 children, at least one preschool child, premarital cohabitation, median values (for women) for the other covariates.}
    \label{fig: Person A}
\end{figure}

\noindent Besides the decomposition of interpretable covariate effects, another strength of a joint model is the option to predict individual probabilities to still be in the relationship after the last measurement of a person. This may be useful for intervention planning in different research questions on the micro-level. In contrast to a classical TVC approach, these predictions are based on the whole estimated longitudinal trajectory of marriage satisfaction and do not only rely on the last measured value. 
Figure \ref{fig: Person A} shows the predictions of a fictional part-time working woman, who lived together with her partner before marriage. She lives with two children, at least one being a preschool child and shows median values (for women) for the other covariates. The only part that varies between the four plots is her trajectory of marital satisfaction indicated by the crosses left to the dashed line showing the time of the last measurement. 
The predicted probabilities are presented with confidence intervals for time-points in the future. Looking at a certain point of time in the future (black solid line), the different trajectories result in different predicted probabilities to still be in the marriage, even though each trajectory ends with the same last measurement of a 4 on the 11-point scale. 
This feature also allows to update the prediction with every newly obtained measurement on a person such that a researcher can trace the development dynamically (see Appendix \ref{append: dynamic_pred}).

\subsection{Comparison to other modelling approaches}
\label{sec: comparison}

In this section, we compare the estimation results (coefficients, standard errors, predictive performance (Mean Squared Error)) of the joint model with two other modelling approaches (TVC approach in a Cox-proportional hazards model and a two-stage model). Comparison tables of the estimation results can be found in Appendix \ref{append: comparison}. 

A classical TVC approach in a time-to-event model\footnote{A classical TVC approach with a lagged satisfaction value (value of previous interview) did not reveal large differences in terms of the size of the coefficient (tendency to smaller effect size) and inference compared to the non-lagged TVC approach.} would heavily underestimate the effect of marital satisfaction on the risk of marriage dissolution (e.g.\ women: TVC approach: $-0.3083$, joint model: $-0.5552$).
This may result from the fact that the risk of marriage dissolution is not only dependent on the observed current value but due to the estimation routine on the whole trajectory of the marriage satisfaction until the point in time.
Modelling marital satisfaction as an endogenous covariate results in some differences regarding the other covariates in the time-to-event submodel as well. For example, the effect of the division of household work for women on the risk of marriage dissolution is overestimated in a classical TVC model and the standard error is underestimated which leads to a smaller p-value.
These differences highlight the importance of the correct model choice when dealing with endogenous covariates in time-to-event models.

In our example the differences between a two-stage model and a joint model are only small and the tendency to underestimate the uncertainty of the satisfaction variable can only be found in the model for women. The regression coefficients for marital satisfaction differ only slightly with $-0.5347$ in the two-stage model and $-0.5552$ in the joint model, and the standard error of $0.0530$ in the two-stage model is smaller than the standard error of $0.0551$ of the presented joint model (men: two-stage: $-0.4240$, s.e.$= 0.0708$; joint model: $-0.4534$, s.e.$= 0.0702$).

Focussing on the longitudinal submodel, we compare the standalone longitudinal model (mixed model fitted with \texttt{lme()}) and the longitudinal submodel of a joint model which controls for the non-random drop-out due to marriage dissolution. There are only small differences in coefficients and standard errors between the two models. However, in the joint model for women the effect of time of marriage has a more pronounced U-shape i.e.\ larger absolute coefficients for the linear and quadratic term.
In the model for men, modelling the non-random drop-out by marriage dissolution leads to a change in the size and significance of the coefficient of premarital cohabitation.
Note that the differences between the longitudinal model approaches may be larger in other applications when a larger share of events (higher number of drop-outs, i.e.\ more marriage dissolutions) is present.

We further evaluated the predictive performance of the three modelling approaches via predicting the event probability for several time points after the last individual measurement of marital satisfaction. Figure \ref{fig:prediction_comparison} shows the mean squared error (MSE) of the models at specific points in time and highlights the advantages of a joint model regarding predictions in the longrun. While the MSE is smallest for the TVC approach when predicting the event outcome up until six months after the last measurement, the MSE for the joint model is smaller when looking at later times and outperforms the two competing models after ten months.

\section{Summary \& Conclusion}

This tutorial paper aimed to introduce the method of a joint model for longitudinal and time-to-event data in the field of social science research. We demonstrate the suitability and added value of answering research questions with endogenous covariates in an application on marital satisfaction and marriage dissolution. Based on the pairfam data, our results indicate that the effect of marital satisfaction on the risk of marriage dissolution is larger than a time-varying Cox model suggests. The strength of the decomposition of effects has been demonstrated and shows e.g.\ that the relative load of household work in a marriage has no direct effect on the risk of marital dissolution for women but for men and a strong indirect effect via marital satisfaction for both sexes.
We believe that this model class is a useful tool in social science research and hope to contribute to its increasing usage.


For the sake of illustration, this tutorial paper kept the modelling structure as simple as possible such that several extensions of the same data example may arise. From a modelling point of view, different association structures may be useful. We therefore tested the joint model with different association structures (current slope, current value \emph{and} current slope, cumulative effect, lagged effect, see Appendix \ref{append: assoc structures}). Model choice criteria suggest to favor the current value association with a lag over the other association structures for men whereas the current value \emph{and} current slope association is the favored model for women. In terms of model set-up, the variable of marital satisfaction is not perfectly normally distributed and another outcome distribution modelled via a Generalized Linear Mixed Model may be a more appropriate choice for the longitudinal submodel. This can be included using the \texttt{JMbayes2} package \citep{JMbayes2} in \texttt{R}. Furthermore, the individual-specific effect of time in the LMM could be modelled non-linearly with semi-parametric methods via splines. Combining statistical modelling with machine learning methods for variable selection may be useful. A model-based boosting algorithm for joint models has been developed by \cite{Griesbach2023_jmboost}.

Regarding the content level, several extensions are conceivable: In order to exploit the data richness of pairfam even further, couple-level data analysis might give additional insights \citep{Ruppanner2017, Hickey2016}. This is a strong limitation of the performed analysis, since the occurrence of marriage dissolution and its timing are assumed to be influenced only by the satisfaction level of one partner in the marriage. The predictive performance might be improved using variables of the partner in the model as well. In addition, one could rethink the exogeneity assumption of the other time-varying covariates in the time-to-event model and also model them as endogeneous in a joint model, e.g.\ regarding the potential anticipation effect of divorce and its connection to working behaviour for women \citep{poortman2005}.

\paragraph{Acknowledgments:}
This paper uses data from the German Family Panel pairfam, coordinated by Josef Brüderl, Sonja Drobnič, Karsten Hank, Johannes Huinink, Bernhard Nauck, Franz J. Neyer, and Sabine Walper. The study was funded from 2004 to 2022 as a priority program and long-term project by the German Research Foundation (DFG).

\noindent The work on this article was supported by the DFG (Number 426493614) and the Volkswagen Foundation (Freigeist Fellowship).

\pagebreak
\bibliographystyle{apalike}  
\bibliography{ref}

\newpage
\appendix

\section{Appendix}

\setcounter{table}{0}
\renewcommand{\thetable}{\Alph{section}.\arabic{table}}

\setcounter{figure}{0}
\renewcommand{\thefigure}{\Alph{section}.\arabic{figure}}

\subsection{Relative load of household work: Index building}
\label{append: index_building household}

The questions were presented to respondents with a partner with whom they live. Five items were summed up, divided by the number of tasks which the couple shares and the mean of the scale (3) was substracted.

\vspace{0.3 cm}

\noindent "I would now like to ask you about how you and your partner organize your daily lives. To what extent do you and [name of current partner (hpn)] share duties in the following domains? If you have a housemaid, nanny, or similar household help, then refer in your answers only to the portion of the work done by you and/or your partner."
\vspace{0.3 cm}
\begin{itemize}
    \item Housework (washing, cooking, cleaning)
    \item Shopping
    \item Working on the house, apartment, or car
    \item Financial and administrative matters
    \item Respondents with children in household: Taking care of the children
\end{itemize}
\vspace{0.3 cm}
Who takes care of that?
\vspace{0.3 cm}
\begin{enumerate}[label=\arabic*]
    \item (Almost) completely my partner
    \item For the most part my partner
    \item Split about 50/50
    \item For the most part me
    \item (Almost) completely me 
    \item []
    \item Only another person
    \item Doesn’t apply to our situation
\end{enumerate}
\noindent Zero was assigned to "Doesn’t apply to our situation" and "Only another person"
The number of tasks with given answers of 1 to 5 is identified and used as denominator. Entries with less than 4 valid responses, i.e.\ less than 4 tasks shared by the couple are coded to missing. 
Substracting 3, centers the variable to 0 for equally divided housework. Negative values indicate less housework than partner and positive more responsibilities in the housework than the partner. 

\subsection{Gender role attitudes: Index building}
\label{append: index_building gender}

In order to control for gender role attitudes, three items were used to built an index. The 5 point scale reached from 1 "disagree completely" to 5 "agree completely".
\vspace{0.3 cm}

"Please tell me how strongly you personally agree with the following statements."
\vspace{0.3 cm}

\begin{itemize}
    \item Women should be more concerned about their family than about their career.
    \item Men should participate in housework to the same extent as women.
    \item A child under 6 will suffer from having a working mother.
\end{itemize}
\vspace{0.3 cm}
The scale of item one and three has been reversed, and the responses to the three items were added and centered around the mean of the summed scale (9). Values above zero indicate liberal gender role attitudes, negative values refer to conservative attitudes.

\newpage
\subsection{Model comparison}
\label{append: comparison}

\subsubsection{Longitudinal model}

\begin{table}[ht]
\caption{Model comparison table for \emph{women}: Longitudinal (sub)model for modelling marital satisfaction.}
\centering
\resizebox{\textwidth}{!}{
\begin{tabular}{lrrr|rrr}
  \hline
       & \multicolumn{3}{c}{Linear mixed model } &  \multicolumn{3}{c}{Longitudinal submodel}\\
  \vspace{0.1cm}
     & \multicolumn{3}{c}{\texttt{lme()}} &  \multicolumn{3}{c}{\texttt{JM()}}\\
     \vspace{0.1cm}
  Variable & Estimate & Std.\ err.  & p-value & Estimate & Std.\ err. & p-value \\ 
  \hline
(Intercept) & 8.7180 & 0.1098 & 0.0000 & 8.7403 & 0.1054 & 0.0000 \\ 
  Time & -2.8256 & 0.4079 & 0.0000 & -3.0387 & 0.4023 & 0.0000 \\ 
  Time$^2$ & 1.8383 & 0.5152 & 0.0004 & 2.0987 & 0.5022 & 0.0000 \\ 
  Relative load of household work & -0.1402 & 0.0192 & 0.0000 & -0.1369 & 0.0191 & 0.0000 \\ 
  Premarital cohabitation$^a$: yes & -0.0988 & 0.0861 & 0.2516 & -0.1036 & 0.0788 & 0.1885 \\ 
  Age at marriage & -0.1915 & 0.0387 & 0.0000 & -0.1820 & 0.0361 & 0.0000 \\ 
  Preschool child(ren) in hh$^a$: yes & -0.0599 & 0.0418 & 0.1517 & -0.0693 & 0.0412 & 0.0924 \\ 
  Number of children in hh$^b$: 1 & -0.2696 & 0.0669 & 0.0001 & -0.2520 & 0.0648 & 0.0001 \\ 
  Number of children in hh$^b$: 2 & -0.2752 & 0.0763 & 0.0003 & -0.2594 & 0.0729 & 0.0004 \\ 
  Number of children in hh$^b$: more & -0.2579 & 0.0970 & 0.0079 & -0.2182 & 0.0921 & 0.0178 \\ 
  Years of education & 0.0868 & 0.0337 & 0.0100 & 0.0672 & 0.0310 & 0.0299 \\ 
  Personal net income & -0.0469 & 0.0249 & 0.0595 & -0.0457 & 0.0248 & 0.0651 \\ 
  Relationship duration at marriage & 0.0328 & 0.0359 & 0.3612 & 0.0183 & 0.0340 & 0.5910 \\ 
  Labor force status$^c$: not working & 0.0314 & 0.0642 & 0.6245 & 0.0474 & 0.0625 & 0.4479 \\ 
  Labor force status$^c$: other & -0.0683 & 0.0696 & 0.3264 & -0.0577 & 0.0677 & 0.3941 \\ 
  Labor force status$^c$: part-time employed & -0.0288 & 0.0570 & 0.6128 & -0.0170 & 0.0555 & 0.7594 \\ 
  Gender role attitudes & -0.0049 & 0.0222 & 0.8265 & -0.0082 & 0.0219 & 0.7069 \\
   \hline
\end{tabular}
}
\begin{FlushLeft}
\small{Reference categories:
$^a$ no, 
$^b$ zero,
$^c$ full-time employed
}
\end{FlushLeft}
\end{table}

\begin{table}[ht]
\centering
\caption{Model comparison table for \emph{men}: Longitudinal (sub)model for modelling marital satisfaction.}
\resizebox{\textwidth}{!}{
\begin{tabular}{lrrr|rrr}
  \hline
       & \multicolumn{3}{c}{Linear mixed model } &  \multicolumn{3}{c}{Longitudinal submodel}\\
  \vspace{0.1cm}
     & \multicolumn{3}{c}{\texttt{lme()}} &  \multicolumn{3}{c}{\texttt{JM()}}\\
     \vspace{0.1cm}
  Variable & Estimate & Std. err.  & p-value & Estimate & Std. err. & p-value \\ 
  \hline
(Intercept) & 8.9591 & 0.1221 & 0.0000 & 9.0393 & 0.1233 & 0.0000 \\ 
  Time & -2.7153 & 0.4870 & 0.0000 & -2.8529 & 0.4976 & 0.0000 \\ 
  Time$^2$ & 1.5466 & 0.6785 & 0.0227 & 1.5018 & 0.7095 & 0.0343 \\ 
  Relative load of household work & -0.0682 & 0.0267 & 0.0106 & -0.0603 & 0.0267 & 0.0237 \\ 
  Premarital cohabitation$^a$: yes & -0.1696 & 0.1037 & 0.1019 & -0.2385 & 0.1052 & 0.0234 \\ 
  Age at marriage & -0.1569 & 0.0433 & 0.0003 & -0.1686 & 0.0417 & 0.0001 \\ 
  Preschool child(ren) in hh$^a$: yes & 0.0273 & 0.0501 & 0.5858 & 0.0249 & 0.0507 & 0.6238 \\ 
  Number of children in hh$^b$: 1 & -0.3394 & 0.0785 & 0.0000 & -0.3302 & 0.0790 & 0.0000 \\ 
  Number of children in hh$^b$: 2 & -0.4088 & 0.0889 & 0.0000 & -0.3910 & 0.0901 & 0.0000 \\ 
  Number of children in hh$^b$: more & -0.3674 & 0.1138 & 0.0012 & -0.3562 & 0.1152 & 0.0020 \\ 
  Years of education & 0.0351 & 0.0381 & 0.3570 & 0.0341 & 0.0393 & 0.3864 \\ 
  Personal net income & -0.0021 & 0.0192 & 0.9137 & 0.0017 & 0.0192 & 0.9307 \\ 
  Relationship duration at marriage & -0.0118 & 0.0383 & 0.7592 & -0.0015 & 0.0420 & 0.9708 \\ 
  Labor force status$^c$: not working & -0.2637 & 0.0970 & 0.0066 & -0.2574 & 0.0977 & 0.0085 \\ 
  Labor force status$^c$: other & -0.3112 & 0.0894 & 0.0005 & -0.3296 & 0.0927 & 0.0004 \\ 
  Labor force status$^c$: part-time employed & -0.2695 & 0.1233 & 0.0289 & -0.2288 & 0.1191 & 0.0548 \\ 
  Gender role attitudes & 0.0253 & 0.0260 & 0.3315 & 0.0281 & 0.0263 & 0.2846 \\
   \hline
\end{tabular}
}
\begin{FlushLeft}
\small{Reference categories:
$^a$ no, 
$^b$ zero,
$^c$ full-time employed
}
\end{FlushLeft} 
\end{table}

\newpage
\subsubsection{Time-to-event model}

\begin{sidewaystable}[pht!]
\centering
\caption{Model comparison table for \emph{women}: Time-to-event (sub)model for modelling the risk of marriage dissolution.} 
\begin{tabular}{lrrr|rrr|rrr}
  \hline
       & \multicolumn{3}{c}{Time-varying covariate} &  \multicolumn{3}{c}{Two-stage model}&  \multicolumn{3}{c}{Time-to-event submodel}\\
  \vspace{0.1cm}
     & \multicolumn{3}{c}{\texttt{coxph()}} &  \multicolumn{3}{c}{\texttt{lme()} and \texttt{coxph()}}&  \multicolumn{3}{c}{\texttt{JM()}}\\
     \vspace{0.1cm}
  Variable & Estimate & Std. err.  & p-value & Estimate & Std. err. & p-value& Estimate & Std. err. & p-value \\ 
  \hline
Years of education & -0.2420 & 0.0968 & 0.0090 & -0.2346 & 0.0957 & 0.0097 & -0.2228 & 0.0969 & 0.0214 \\ 
  Age at marriage & -0.0788 & 0.1035 & 0.4418 & -0.1413 & 0.1060 & 0.1909 & -0.1546 & 0.1070 & 0.1483 \\ 
  Preschool child(ren) in hh$^a$: yes & -0.3780 & 0.2096 & 0.0773 & -0.3620 & 0.2115 & 0.0975 & -0.3286 & 0.2121 & 0.1213 \\ 
  Number of children in hh$^b$: 1 & 0.0463 & 0.2931 & 0.8713 & 0.0451 & 0.2931 & 0.8734 & 0.0893 & 0.2969 & 0.7637 \\ 
  Number of children in hh$^b$: 2 & 0.1254 & 0.3017 & 0.6803 & 0.0411 & 0.3046 & 0.8939 & 0.0508 & 0.3082 & 0.8691 \\ 
  Number of children in hh$^b$: more & 0.1319 & 0.3576 & 0.7140 & 0.0804 & 0.3619 & 0.8258 & 0.0799 & 0.3649 & 0.8266 \\ 
  Relative load of household work & 0.1451 & 0.0771 & 0.0713 & 0.0715 & 0.0796 & 0.3968 & 0.1089 & 0.0793 & 0.1694 \\ 
  Premarital cohabitation$^a$: yes & 0.0526 & 0.2197 & 0.8182 & 0.1019 & 0.2189 & 0.6586 & 0.1297 & 0.2225 & 0.5599 \\ 
  Personal net income & -0.0283 & 0.1470 & 0.7439 & -0.0328 & 0.1600 & 0.7509 & -0.0230 & 0.1462 & 0.8750 \\ 
  Relationship duration at marriage & -0.2673 & 0.1025 & 0.0080 & -0.2512 & 0.1035 & 0.0149 & -0.2570 & 0.1047 & 0.0140 \\ 
  Labor force status$^c$: not working & -0.3843 & 0.3039 & 0.2014 & -0.3509 & 0.3085 & 0.2475 & -0.3514 & 0.3051 & 0.2494 \\ 
  Labor force status$^c$: other & -0.0363 & 0.2820 & 0.8983 & -0.0545 & 0.2833 & 0.8491 & -0.0238 & 0.2830 & 0.9329 \\ 
  Labor force status$^c$: part-time employed & -0.1260 & 0.2403 & 0.6037 & -0.1647 & 0.2409 & 0.4961 & -0.1381 & 0.2422 & 0.5685 \\ 
  Gender role attitudes & 0.0848 & 0.0887 & 0.3245 & 0.0876 & 0.0892 & 0.3214 & 0.0819 & 0.0896 & 0.3606 \\ 
  Satisfaction & -0.3083 & 0.0253 & 0.0000 & -0.5347 & 0.0530 & 0.0000& -0.5552 & 0.0551 & 0.0000\\ 
   \hline
\end{tabular}
\begin{FlushLeft}
\small{Reference categories:
$^a$ no, 
$^b$ zero,
$^c$ full-time employed
}
\end{FlushLeft} 
\end{sidewaystable}


\begin{sidewaystable}[pht!]
\centering
\caption{Model comparison table for \emph{men}: Time-to-event (sub)model for modelling the risk of marriage dissolution.} 
\begin{tabular}{lrrr|rrr|rrr}
  \hline
       & \multicolumn{3}{c}{Time-varying covariate} &  \multicolumn{3}{c}{Two-stage model}&  \multicolumn{3}{c}{Time-to-event submodel}\\
  \vspace{0.1cm}
     & \multicolumn{3}{c}{\texttt{coxph()}} &  \multicolumn{3}{c}{\texttt{lme()} and \texttt{coxph()}}&  \multicolumn{3}{c}{\texttt{JM()}}\\
     \vspace{0.1cm}
  Variable & Estimate & Std. err.  & p-value & Estimate & Std. err. & p-value& Estimate & Std. err. & p-value \\ 
  \hline
Years of education & -0.1012 & 0.1228 & 0.4061 & -0.1292 & 0.1234 & 0.2961 & -0.1266 & 0.1232 & 0.3043 \\ 
  Age at marriage & 0.1366 & 0.1384 & 0.2970 & 0.1255 & 0.1378 & 0.3418 & 0.1291 & 0.1383 & 0.3508 \\ 
  Preschool child(ren) in hh$^a$: yes & 0.0164 & 0.2809 & 0.9573 & 0.0064 & 0.2794 & 0.9832 & 0.0196 & 0.2806 & 0.9442 \\ 
  Number of children in hh$^b$: 1 & -0.1469 & 0.4026 & 0.7291 & -0.1767 & 0.4011 & 0.6667 & -0.1158 & 0.4041 & 0.7744 \\ 
  Number of children in hh$^b$: 2& 0.3322 & 0.3800 & 0.3978 & 0.2857 & 0.3781 & 0.4570 & 0.3430 & 0.3825 & 0.3699 \\ 
  Number of children in hh$^b$: more & 0.3797 & 0.4603 & 0.4021 & 0.3476 & 0.4585 & 0.4335 & 0.3927 & 0.4633 & 0.3967 \\ 
  Relative load of household work & 0.2502 & 0.1294 & 0.0206 & 0.2305 & 0.1311 & 0.0343 & 0.2413 & 0.1312 & 0.0660 \\ 
  Premarital cohabitation$^a$: yes & -0.3325 & 0.3001 & 0.2974 & -0.3453 & 0.3003 & 0.2784 & -0.3357 & 0.3020 & 0.2663 \\ 
  Personal net income & -0.0552 & 0.1679 & 0.7828 & -0.0710 & 0.1762 & 0.7462 & -0.0481 & 0.1647 & 0.7703 \\ 
  Relationship duration at marriage & -0.0722 & 0.1194 & 0.6172 & -0.0704 & 0.1201 & 0.6286 & -0.0709 & 0.1208 & 0.5573 \\ 
  Labor force status$^c$: not working & -0.1215 & 0.4968 & 0.8130 & -0.0195 & 0.4982 & 0.9697 & 0.0613 & 0.4931 & 0.9012 \\ 
  Labor force status$^c$: other & -0.0453 & 0.3583 & 0.8939 & -0.0528 & 0.3587 & 0.8764 & -0.0242 & 0.3599 & 0.9465 \\ 
  Labor force status$^c$: part-time employedt & -1.3864 & 1.0184 & 0.1806 & -1.2501 & 1.0177 & 0.2260 & -1.1292 & 1.0137 & 0.2653 \\ 
  Gender role attitudes & 0.1173 & 0.1167 & 0.3469 & 0.1288 & 0.1162 & 0.2985 & 0.1294 & 0.1162 & 0.2654 \\ 
  Satisfaction & -0.2798 & 0.0351 & 0.0000  & -0.4240 & 0.0708 & 0.0000& -0.4534 & 0.0702 & 0.0000\\ 
   \hline
\end{tabular}
\begin{FlushLeft}
\small{Reference categories:
$^a$ no, 
$^b$ zero,
$^c$ full-time employed
}
\end{FlushLeft}
\end{sidewaystable}

\newpage

\subsection{Predictive Performance}
\label{append: predperf}
\begin{figure}[h]
    \centering
    \includegraphics[width=\textwidth]{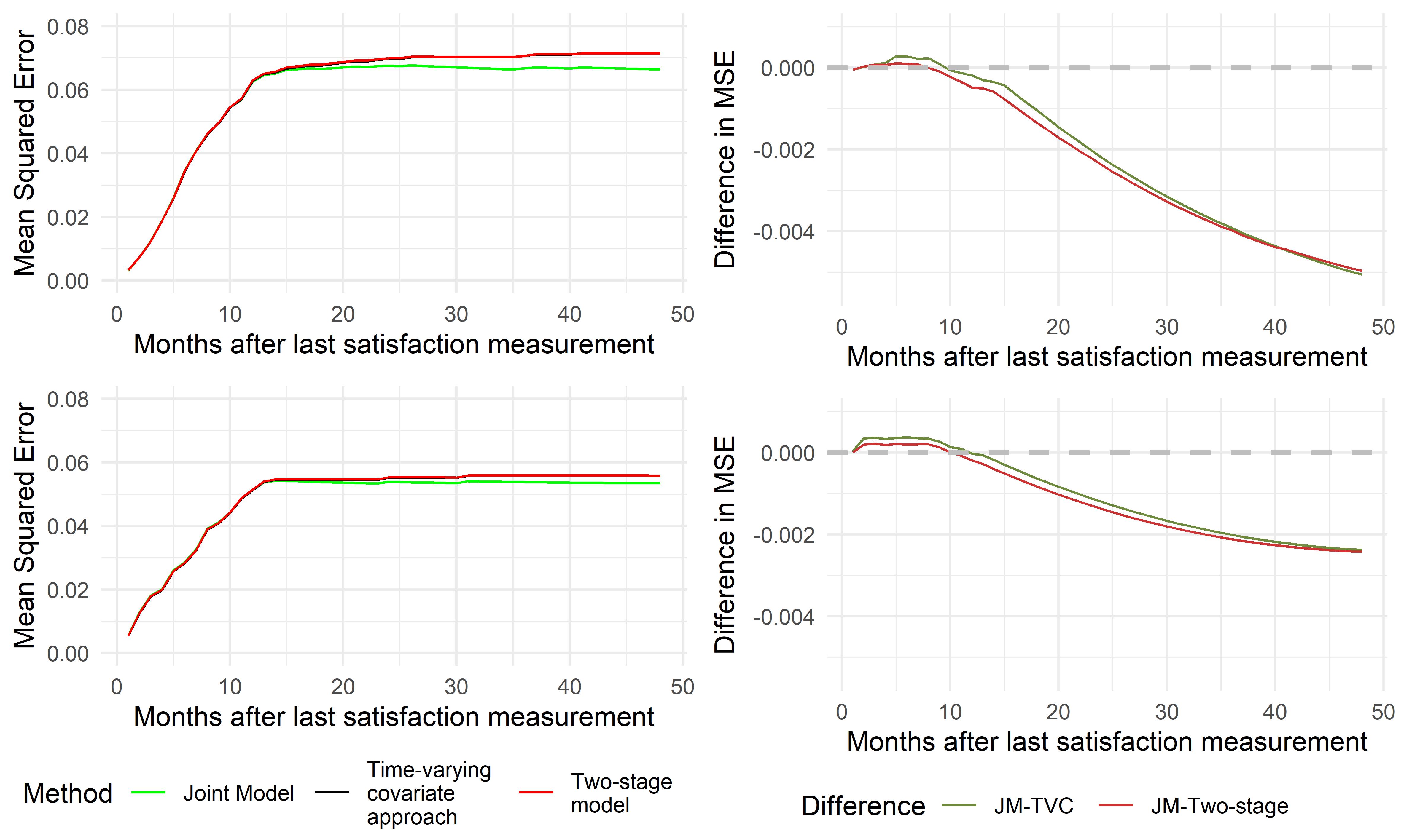}
    \caption{Upper panel: Women, Lower panel: Men. \\ Left: Comparison of Mean Squared Error (MSE) of three different model approaches by time since last measurement. Right: Difference in Joint Model vs.\ Cox model and Difference in Joint Model vs.\ Two-stage model in terms of MSE by time since last measurement.}
    \label{fig:prediction_comparison}
\end{figure}

\newpage

\subsection{Different association structures}
\label{append: assoc structures}
The model for female and male respondents has been tested with different association structures as well. The four different association structures can be characterised as follows:

\begin{itemize}
    \item Current value: 
    \item[] $        h(t|M_i(t), \boldsymbol{x_i}) = h_0(t) \exp [\boldsymbol \gamma^T \boldsymbol{{x}_i}_{\text{surv}} + \alpha m_i(t)] $
    \item Current value and current slope: 
    \item[] $        h(t|M_i(t), \boldsymbol{x_i}) = h_0(t) \exp [\boldsymbol \gamma^T \boldsymbol{{x}_i}_{\text{surv}} + \alpha_1 m_i(t)+\alpha_2 m'_i(t)]$
    \item Cumulative effect (area under longitudinal trajectory): 
    \item[] $h(t|M_i(t), \boldsymbol{x_i}) = h_0(t) \exp [\boldsymbol \gamma^T \boldsymbol{{x}_i}_{\text{surv}} + \alpha 
 \int_0^t m_i(s) ds]$
 \item Lagged effect, where $c$ defines the desired time lag: 
 \item[] $h(t|M_i(t), \boldsymbol{x_i}) = h_0(t) \exp [\boldsymbol \gamma^T \boldsymbol{{x}_i}_{\text{surv}} + \alpha 
 m_i\{\text{max}(t-c),0\}]$ 
\end{itemize}
The respective model choice criteria values for the models with different association structures are given in the table below.
\begin{table}[ht!]
\centering
\begin{tabular}{rcc}
  \hline
 & Women  & Men \\ 
  \hline
Current value & 66934.06 & 45255.43 \\ 
  Current slope & 67004.06 & 45277.44 \\ 
  Current value+current slope & \textbf{66933.88} & 45258.63 \\ 
  Cumulative & 66956.5 & 45267.12 \\ 
 Current value with lag (one month) & 66934.61 & \textbf{45253.71} \\ 
   \hline
\end{tabular}
\caption{AIC for joint models with different association structures. }
\end{table}
\newpage

\subsection{Dynamic prediction}
\label{append: dynamic_pred}

\begin{figure}[h!]
    \centering
    \includegraphics[width=\textwidth]{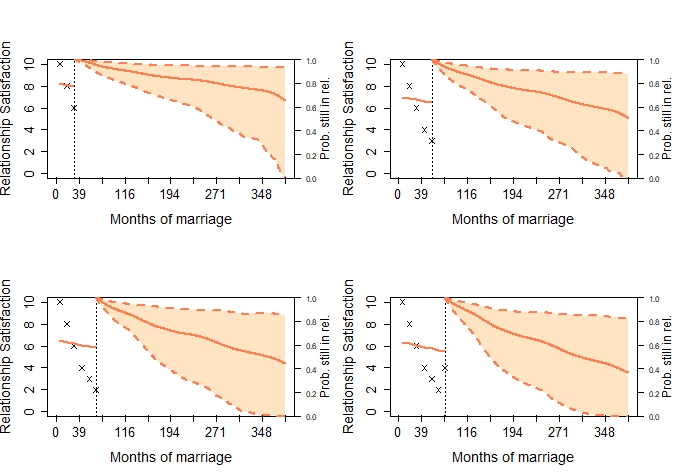}
    \caption{Predicted survival probability for a person updating her marital satisfaction trajectory. Covariate values: woman, part-time working, 2 children, at least one preschool child, premarital cohabitation, median values (for women) for the other covariates.}
    \label{fig: Person A updated}
\end{figure}

Consider the upper right panel of Figure \ref{fig: Person A}: Each new observation of marital satisfaction leads to an updated individual predicted probability to still be in the marriage as depicted in Figure \ref{fig: Person A updated}.

\end{document}